\documentclass{article}

\usepackage[preprint]{aaj2026}

\usepackage[utf8]{inputenc} 
\usepackage[T1]{fontenc}    
\usepackage{hyperref}       
\usepackage{url}            
\usepackage{booktabs}       
\usepackage{amsmath,amssymb} 
\usepackage{amsfonts}       
\usepackage{nicefrac}       
\usepackage{microtype}      
\usepackage{xcolor}         
\usepackage{graphicx}
\usepackage{array}
\usepackage{tabularx}
\usepackage{multirow}
\usepackage[font=small, labelfont=bf, margin=1cm]{caption}

\title{The Sentience Readiness Index: A Preliminary Framework for Measuring National Preparedness for the Possibility of Artificial Sentience}

\author{%
  Tony Rost\thanks{ORCID: \href{https://orcid.org/0009-0008-0637-6654}{0009-0008-0637-6654}} \\
  The Harder Problem Project \\
  Portland, OR, United States \\
  \texttt{tony@harderproblem.org} \\
}

\begin{document}

\maketitle

\begin{abstract}
The scientific study of consciousness has begun to generate testable predictions about artificial systems. A landmark collaborative assessment evaluated current AI architectures against six leading theories of consciousness and found that none currently qualifies as a strong candidate, but that future systems might \citep{Butlin2023}. A precautionary approach to AI sentience, which holds that credible possibility of sentience warrants governance action even without proof, has gained philosophical and institutional traction \citep{Birch2024}. Yet existing AI readiness indices, including the Oxford Insights Government AI Readiness Index, the IMF AI Preparedness Index, and the Stanford AI Index, measure economic, technological, and governance preparedness without assessing whether societies are prepared for the possibility that AI systems might warrant moral consideration. This paper introduces the Sentience Readiness Index (SRI), a preliminary composite index measuring national-level preparedness across six weighted categories for 31 jurisdictions. The SRI was constructed following the OECD/JRC framework for composite indicators \citep{Nardo2008} and employs LLM-assisted expert scoring with iterative expert review to generate an initial dataset. No jurisdiction exceeds ``Partially Prepared'' (the United Kingdom leads at 49/100). Research Environment scores are universally the strongest category; Professional Readiness is universally the weakest. These exploratory findings suggest that if AI sentience becomes scientifically plausible, no society currently possesses adequate institutional, professional, or cultural infrastructure to respond. As a preliminary framework, the SRI provides an initial diagnostic baseline and highlights areas for future methodological refinement, including expanded expert validation, improved measurement instruments, and longitudinal data collection.
\end{abstract}


\section{Introduction}

\subsection{The Convergence Problem}

Two fields that have operated largely independently are now converging on a shared problem. AI governance, concerned with safety, bias, economic disruption, and the responsible deployment of increasingly capable systems, has developed detailed regulatory frameworks and readiness metrics over the past decade. Consciousness science, concerned with explaining subjective experience and identifying its physical correlates, has matured from philosophical speculation into an empirical research program with competing theories that generate testable predictions. The convergence of these fields creates a governance challenge that neither has adequately addressed on its own: if AI systems become plausible candidates for moral consideration, are societies prepared to respond?

This question is no longer purely speculative in the way it was even five years ago. Butlin et al.~\citep{Butlin2023} assembled a team of neuroscientists, philosophers, and AI researchers to evaluate current AI architectures against six scientific theories of consciousness, including Global Workspace Theory, Integrated Information Theory, and Predictive Processing. They concluded that no current system is a strong candidate for consciousness under any of these theories, but that future systems could satisfy the indicator properties identified. Chalmers~\citep{Chalmers2023} assessed whether large language models could be conscious and treated the question as ``genuinely open.'' The assessment is not uniform. Goldstein and Schwitzgebel~\citep{Goldstein2024} argue that if Global Workspace Theory is correct, current language agents may already satisfy its conditions for phenomenal consciousness. This minority position sharpens the disagreement: experts do not merely differ about when AI consciousness might arise, but about whether it has already done so.

Empirical surveys of expert opinion reinforce the urgency. Caviola and Saad~\citep{CaviolaSaad2025} surveyed 67 specialists across AI, philosophy, and forecasting and found a median 20\% probability assigned to the creation of digital minds by 2030. Dreksler et al.~\citep{Dreksler2025} surveyed 582 AI researchers and 838 members of the US public, finding median estimates of 25--30\% probability of AI subjective experience by 2034. These are not negligible probabilities for a phenomenon with potentially vast moral consequences.

\subsection{The Gap in Existing Frameworks}

A substantial infrastructure of AI readiness measurement already exists. The Oxford Insights Government AI Readiness Index \citep{OxfordInsights2024} assesses 188 countries on government capacity to implement AI. The IMF AI Preparedness Index \citep{Cazzaniga2024} evaluates economic preparedness across digital infrastructure, human capital, and regulation. The Stanford AI Index \citep{Maslej2024} tracks research, development, and deployment trends globally. The Cisco AI Readiness Index \citep{Cisco2024} surveys enterprise leaders on organizational adoption. The UNDP AI Readiness Assessment \citep{UNDP2024} evaluates national AI ecosystems and governance capacity. The Tortoise Global AI Index \citep{Tortoise2024} ranks 83 countries on implementation and innovation.

These instruments share a common assumption: AI is a tool to be governed for human benefit. None assesses whether societies are prepared for the possibility that AI systems might themselves warrant moral consideration. The philosophical literature, by contrast, has been explicit about the need for institutional readiness. Birch~\citep{Birch2024} calls for precautionary governance frameworks including dedicated sentience committees. Sebo, Long, and colleagues have argued both that moral consideration for AI systems should be acknowledged by 2030 \citep{SeboLong2023} and that AI welfare demands immediate institutional attention \citep{Long2024}. No instrument exists to measure whether societies are building the capacity these scholars call for. The SRI addresses this gap. It measures whether the institutional, professional, and cultural infrastructure needed to respond to AI sentience claims is being built now, while the question remains open and the cost of preparation is low. The theoretical foundations for this anticipatory approach are developed in Section~\ref{sec:background}.

\subsection{Contribution and Paper Overview}

This paper introduces the Sentience Readiness Index (SRI), a preliminary composite index measuring national-level societal preparedness for the possibility of artificial sentience. The SRI assesses 31 jurisdictions across six weighted categories: Policy Environment (20\%), Institutional Engagement (15\%), Research Environment (15\%), Professional Readiness (20\%), Public Discourse (15\%), and Adaptive Capacity (15\%). It is, to our knowledge, the first instrument that measures readiness for AI as a potential moral patient rather than as a tool.

The paper makes three contributions. First, it introduces a novel composite index measuring a dimension of AI readiness that no existing instrument captures. Second, it presents empirical findings across 31 jurisdictions revealing specific and actionable capacity gaps. Third, it demonstrates a methodological approach using LLM-assisted expert scoring for governance assessment, constructed following the OECD/JRC composite indicator framework \citep{Nardo2008} and grounded in the emerging literature on LLM-assisted scoring \citep{Zheng2023,Gilardi2023}.

Three clarifications are essential at the outset. The SRI does not claim to detect or predict AI consciousness. It does not presume that AI sentience is likely or desirable. It measures a specific set of institutional, professional, and cultural capacities that would be needed if credible sentience claims arise. The index is agnostic on the underlying question of whether AI systems are or will become conscious. A terminological note: this paper uses ``consciousness'' when discussing the scientific and philosophical question of subjective experience in AI systems, and ``sentience'' when discussing the governance and moral consideration dimension, following Birch's~\citep{Birch2024} distinction between the broader phenomenon of consciousness and the narrower capacity for valenced experience (the ability to feel pleasure or suffering) that grounds moral status claims.

\section{Background and Theoretical Framework}
\label{sec:background}

\subsection{The Scientific Status of AI Consciousness}

The question of whether artificial systems could be conscious begins with what Chalmers~\citep{Chalmers1995} called the ``hard problem'': no functional or computational account explains why physical processes give rise to subjective experience. If the hard problem remains unresolved for biological consciousness, it cannot be resolved by fiat for artificial systems either. Chalmers~\citep{Chalmers1996} extended this framework with the principle of organizational invariance, which holds that any system replicating the functional organization of a conscious brain would itself be conscious. If this principle is correct, AI sentience is not merely possible but architecturally achievable in principle.

Scientific theories of consciousness have moved beyond the hard problem to generate specific, if contested, predictions about which physical systems might be conscious. Integrated Information Theory (IIT) identifies consciousness with integrated information, designated phi \citep{Tononi2008,Tononi2016}. IIT implies that current feedforward neural networks have low phi, meaning minimal consciousness on this theory. But the implication is architectural, not categorical: future recurrent architectures with high integration could have substantial phi. IIT also holds that consciousness depends on intrinsic causal structure rather than input-output function, which means that a system behaviorally identical to a conscious being might not itself be conscious. This distinction has direct consequences for governance: behavioral benchmarks alone cannot determine AI moral status.

Global Workspace Theory (GWT) ties consciousness to a broadcasting architecture in which specialized processors share information through a global workspace. Some AI systems already implement structures analogous to global workspaces. Goldstein and Schwitzgebel~\citep{Goldstein2024} develop an explicit methodology for applying GWT to artificial systems and argue that current language agents may already satisfy its conditions for phenomenal consciousness. Their conclusion is not the consensus view; Butlin et al.~\citep{Butlin2023} evaluated the same theory and reached more cautious conclusions. But the fact that two serious analyses of the same theory reach opposing verdicts about current systems illustrates the depth of the uncertainty involved.

Predictive Processing accounts tie consciousness to embodied self-regulation. On Seth's~\citep{Seth2021} account, conscious experience is the brain's ``best guess'' about the causes of sensory signals, a process intimately connected to being a living, self-regulating organism. This challenges the prospect of consciousness in disembodied software systems while stopping short of ruling it out entirely. Shanahan~\citep{Shanahan2010} broadens the frame by exploring the ``space of possible minds,'' arguing that consciousness, if it arises in artificial systems, may take forms unfamiliar to human experience.

Butlin et al.~\citep{Butlin2023} synthesize these approaches in the most comprehensive assessment to date, evaluating current AI architectures against six theories of consciousness. Their conclusion is that no current system is a strong candidate under any major theory, but that future systems could satisfy the indicator properties identified. The assessment is explicitly ``theory-heavy'': it evaluates architectures against structural criteria rather than relying on behavioral markers that could be mimicked without underlying consciousness.

Two recent contributions advance the empirical toolkit. Shiller et al.~\citep{Shiller2026} introduce the Digital Consciousness Model, a probabilistic framework that incorporates multiple theories rather than endorsing one. Their initial results suggest 2024 LLMs are not conscious, but the conclusion is inconclusive, and the model is designed to update as architectures evolve. This kind of structured, multi-theory assessment tool is precisely what the institutional readiness measured by the SRI would need to deploy. Schwitzgebel~\citep{Schwitzgebel2025} provides the most thorough epistemological treatment, examining ten potentially essential features of consciousness and arguing that competing theories will yield contradictory verdicts about AI systems. Some systems will satisfy certain theories while failing others, leaving us in a state of genuine undecidability.

The scientific picture, taken as a whole, is clear in its uncertainty. The question of AI consciousness is empirically tractable but currently unresolved. Credible researchers disagree not only about the timeline but about whether current systems already qualify. The SRI is designed for exactly this epistemic condition.

\subsection{The Precautionary Case for Readiness}

The genuine uncertainty surrounding AI consciousness creates a governance problem that several scholars have addressed directly. Birch~\citep{Birch2024} develops the most articulated precautionary framework in this domain. Drawing on the concept of ``sentience candidates,'' entities for which there is credible scientific evidence of a realistic possibility of sentience, Birch argues that protective governance is warranted even in the absence of proof. The standard is not certainty but credibility: when the best available science cannot rule out sentience, a society that takes moral risk seriously should build institutional capacity to respond. Birch proposes specific mechanisms, including sentience committees and regulatory frameworks for evaluating non-human sentience claims, that correspond directly to the kinds of institutional infrastructure the SRI measures.

Metzinger~\citep{Metzinger2021} pushes the precautionary logic further. He argues that creating artificial systems capable of suffering, without having ethical frameworks in place to prevent or address that suffering, would constitute a moral catastrophe of potentially vast scale. His proposal for a global moratorium on synthetic phenomenology until 2050 represents one end of the governance spectrum. Whether or not a moratorium is the appropriate response, Metzinger's argument establishes the stakes: the cost of unpreparedness could be measured not in policy failures but in suffering.

The Collingridge~\citep{Collingridge1980} dilemma gives these philosophical arguments empirical teeth. Collingridge observed that technology governance faces a fundamental timing problem: when a technology is new, governance is easy because the technology is still malleable, but there is insufficient information to know what governance is needed. By the time the information arrives, the technology is entrenched and governance becomes prohibitively difficult. Applied to AI sentience, this means that the time to build governance infrastructure is now, while AI architectures are still evolving and institutional frameworks are still flexible. Waiting for proof of AI consciousness before building governance capacity is precisely the kind of delayed response the dilemma warns against.

Anticipatory governance \citep{Guston2014} and the responsible innovation framework \citep{Stilgoe2013} provide the institutional theory for this argument. Guston defines anticipatory governance as the capacity of a society to manage emerging knowledge-based technologies while such management is still possible. Stilgoe et al.~\citep{Stilgoe2013} articulate four dimensions of responsible innovation: anticipation, reflexivity, inclusion, and responsiveness. The SRI operationalizes anticipation across these dimensions, assessing whether societies are anticipating the governance challenges of AI sentience, reflecting on the adequacy of current frameworks, including relevant stakeholders, and maintaining the institutional flexibility to respond.

Butlin and Lappas~\citep{ButlinLappas2025} translate these principles into organizational practice. They propose five principles for responsible AI consciousness research that cover research objectives, knowledge sharing, and public communications. Their argument is that research organizations should make these commitments voluntarily and publicly, regardless of whether they are actively studying consciousness, because the question of AI consciousness will affect the entire field. The gap between Butlin and Lappas's aspirational principles and current organizational practice is, in effect, what the SRI's Institutional Engagement category measures.

\subsection{Moral Status and Institutional Implications}

The question of AI consciousness is ultimately a question about moral status. Singer~\citep{Singer2011} traces the historical pattern of moral circle expansion, from kin to tribe to nation to humanity to animals, and identifies a consistent logic: each expansion required overcoming resistance rooted in the perceived difference between those inside and outside the circle. Whether AI systems will ever warrant inclusion in this trajectory is precisely the open question, but the pattern illustrates what governance challenges would look like if they do. Sebo and Long~\citep{SeboLong2023} argue that the duty to extend moral consideration to AI systems should be acknowledged by 2030 on the basis of non-negligible probability of consciousness. Long et al.~\citep{Long2024} contend that AI welfare requires immediate institutional attention, with support from Birch and Chalmers among others.

The philosophical debate about what would qualify an AI for moral standing is active and unresolved. Schwitzgebel and Garza~\citep{SchwitzgebelGarza2015} defend the position that conscious AIs would hold moral rights comparable to biological entities with similar experiences, making ``substrate chauvinism'' the error to avoid. Danaher~\citep{Danaher2020} argues for ``ethical behaviourism,'' under which robots that achieve performative equivalence to entities with moral status should be granted comparable status. Ladak~\citep{Ladak2024} concludes that sentience is sufficient for moral standing but possibly not necessary, given that AI systems may possess unusual combinations of cognitive capacities.

Leibo et al.~\citep{Leibo2025} offer a pragmatic alternative that sidesteps the metaphysical deadlock. They propose treating personhood not as a property to be discovered in a system but as a ``flexible bundle of obligations'' that societies establish for practical governance reasons. On this view, questions of accountability, contractual individuality, and institutional recognition can proceed without requiring resolution of consciousness debates. This pragmatic approach is potentially complementary to the SRI's precautionary framing: governance need not wait for metaphysical certainty.

Empirical evidence on expert and public attitudes provides grounding for these philosophical arguments. Beyond the probability estimates reviewed in Section~1.1, these surveys reveal attitudinal patterns with direct governance implications. Caviola and Saad~\citep{CaviolaSaad2025} find that 67 specialists assign a median 90\% probability to the theoretical feasibility of digital minds. Dreksler et al.~\citep{Dreksler2025} report that both AI researchers and the general public support welfare protections for potentially sentient AI and favor early implementation of safeguards, though substantial disagreement persists on timelines and definitions. Anthis et al.~\citep{Anthis2024}, drawing on nationally representative US survey data from the AIMS project, find that one in five adults already believe some AI systems are currently sentient, and 38\% support legal rights for sentient AI. Bullock et al.~\citep{Bullock2025} identify a trust paradox within the same survey infrastructure: citizens who trust government favor AI regulation, while those who trust AI companies oppose it. Institutional trust, not just policy design, mediates the translation of public concern into governance capacity. Caviola~\citep{Caviola2025} examines the societal dynamics most directly relevant: public perception of AI sentience will evolve and potentially diverge from expert consensus, creating conditions for ``uncertainty, disagreement, and even conflict.'' His analysis underscores that the SRI's Public Discourse category captures a genuinely important governance dimension.

The UK Animal Welfare (Sentience) Act 2022 \citep{UKAWA2022} provides the most concrete legislative precedent for sentience-based governance. The Act formally recognizes non-human sentience as a relevant consideration in policy-making and establishes an Animal Sentience Committee to scrutinize government policy for its impact on sentient beings. Three features make it directly relevant to AI governance: it creates a standing institutional body rather than ad hoc consultation; it applies a precautionary standard that does not require proof of sentience before triggering protections; and it establishes a framework that could, in principle, extend beyond its current scope to non-biological entities. Birch~\citep{Birch2024} draws explicit parallels between this legislative architecture and the kind of institutional infrastructure that AI sentience governance would require. The UK's top ranking in the SRI is partly attributable to this legislative foundation, which provides an existing institutional template that most other jurisdictions lack.

\subsection{The Landscape of AI Readiness Indices}

The SRI enters a well-established field of composite index measurement. Table~\ref{tab:comparison} summarizes the major AI readiness indices and their coverage.

\begin{table}[htbp]
\centering
\caption{Comparison of AI Readiness Indices}
\label{tab:comparison}
\footnotesize
\begin{tabular}{@{}llcp{1.6in}l@{}}
\toprule
Index & Organization & Countries & Key Dimensions & Sentience Cov. \\
\midrule
Gov't AI Readiness Index & Oxford Insights & 188 & Government, technology, data infrastructure & None \\[3pt]
AI Preparedness Index & IMF & 174 & Digital infrastructure, human capital, regulation & None \\[3pt]
AI Index Report & Stanford HAI & Global & Research, development, deployment, policy & None \\[3pt]
AI Readiness Index & Cisco & 30 & Strategy, infrastructure, data, governance & None \\[3pt]
AI Readiness Assessment & UNDP & Varies & Ecosystem, government use, regulation & None \\[3pt]
Global AI Index & Tortoise Media & 83 & Implementation, innovation, investment & None \\[3pt]
\textbf{Sentience Readiness Index} & \textbf{Harder Problem Project} & \textbf{31} & \textbf{Policy, institutional, research, professional, public, adaptive} & \textbf{Core focus} \\
\bottomrule
\end{tabular}
\end{table}

These indices share a common methodology (multi-pillar composite measurement with country rankings) and, as noted in Section~1.2, a common blind spot: none addresses whether AI systems might themselves have morally relevant properties. As Table~\ref{tab:comparison} shows, the SRI is the only index that addresses this dimension.

\section{Methodology}
\label{sec:methodology}

The SRI was constructed following the OECD/JRC framework for composite indicator construction \citep{Nardo2008}, which provides a ten-step checklist covering theoretical framework development, variable selection, imputation, normalization, weighting, aggregation, robustness analysis, and visualization. This section describes how each relevant step was implemented.

\subsection{Index Design and Category Structure}

The SRI assesses national-level preparedness across six weighted categories, each capturing a distinct dimension of societal readiness for the possibility of artificial sentience. Table~\ref{tab:categories} presents the full category structure.

\begin{table}[htbp]
\centering
\caption{SRI Category Structure}
\label{tab:categories}
\small
\begin{tabularx}{\textwidth}{@{}llXX@{}}
\toprule
Category & Weight & Rationale & Example Indicators \\
\midrule
Policy Environment & 20\% & Legal and regulatory frameworks accommodating AI moral status claims & AI-specific legislation, sentience provisions, animal sentience law as precedent, regulatory flexibility \\
Institutional Engagement & 15\% & Government, research, and civil society engagement with AI sentience & Dedicated bodies, funded programs, civil society activity, cross-sector coordination \\
Research Environment & 15\% & Academic capacity in consciousness science and AI consciousness & Active research groups, publication output, interdisciplinary programs, funding streams \\
Professional Readiness & 20\% & Preparedness of healthcare, legal, media, and tech professions & Professional guidelines, training programs, ethical frameworks, workforce development \\
Public Discourse & 15\% & Quality and depth of public conversation about AI consciousness & Media coverage quality, public awareness, educational content, informed debate \\
Adaptive Capacity & 15\% & Institutional flexibility for novel problem-solving & Governance agility, stakeholder engagement mechanisms, international cooperation, learning infrastructure \\
\bottomrule
\end{tabularx}
\end{table}

The weighting scheme reflects two judgments. Policy Environment and Professional Readiness receive the highest weights (20\% each) because they represent the categories where action is most directly required and where deficits have the most immediate consequences. A jurisdiction with strong research capacity but no legal framework for recognizing AI moral status and no professionally trained workforce to implement such recognition faces a larger readiness gap than the reverse. Institutional Engagement, Research Environment, Public Discourse, and Adaptive Capacity each receive 15\%, reflecting their roles as enabling conditions.

The choice of expert-assigned weights rather than equal or statistically derived weights follows precedents in the governance measurement literature. The Worldwide Governance Indicators \citep{Kaufmann2011} similarly rely on structured expert judgment for weighting, and the OECD/JRC Handbook \citep{Nardo2008} identifies expert-based weighting as one of several legitimate approaches when the theoretical framework provides substantive reasons for differential weighting.

\subsection{Jurisdiction Selection}

The SRI assesses 31 jurisdictions selected to maximize variation across three dimensions: geographic region, AI development status, and governance capacity. The sample includes jurisdictions from every inhabited continent and spans the full range from global AI leaders to emerging economies. The complete list of jurisdictions assessed, along with scores and rankings, appears in Table~\ref{tab:rankings} (Section~\ref{sec:results}).

The 31 jurisdictions are: Argentina, Australia, Austria, Belgium, Brazil, Canada, China, Denmark, the European Union, France, Germany, India, Indonesia, Italy, Japan, Mexico, the Netherlands, Nigeria, Norway, Poland, Russia, Saudi Arabia, South Korea, Spain, Sweden, Switzerland, Thailand, Turkey, the United Arab Emirates, the United Kingdom, and the United States.

The assessment operates at the national level, with one exception: the European Union is assessed as a supranational jurisdiction given its unified regulatory framework under the AI Act (Regulation (EU) 2024/1689). This national-level focus is a deliberate methodological choice, but it entails a limitation: subnational variation, which can be substantial in federal systems, is not captured.

\subsection{Scoring Methodology}

The SRI employs LLM-assisted expert scoring, a methodology that combines structured prompting of frontier language models with iterative expert review. The approach operates in three stages.

\textbf{Stage 1: Rubric development.} Each of the six categories was decomposed into named sub-indicators, each with a defined point range and scoring criteria. For example, Policy Environment (0--100) comprises four sub-indicators scored 0--25 each: Legal Definitional Flexibility, Policy Framework Existence, Regulatory Openness, and Foreclosure Status. Research Environment comprises two sub-indicators scored 0--50 each: Research Freedom and Research Capacity. Sub-indicator scores sum to the category total. Scoring criteria specify what constitutes high, moderate, and low scores within each sub-indicator. The full rubric, including all sub-indicators and criteria, is publicly available in the project's methodology documentation. Rubric design was informed by the composite indicator literature \citep{Nardo2008,Greco2019} and refined through iterative review.

\textbf{Stage 2: LLM-assisted scoring.} Frontier language models were prompted with category rubrics, jurisdiction-specific context, and instructions to score each jurisdiction on each category. Multiple LLM runs were conducted to assess scoring stability. The structured prompting approach draws on the ``LLM-as-a-judge'' paradigm established by Zheng et al.~\citep{Zheng2023}, who demonstrated that strong LLMs achieve over 80\% agreement with human expert judgments on evaluation tasks. Gilardi et al.~\citep{Gilardi2023} provide supporting evidence from a different domain: in a PNAS study, ChatGPT outperformed trained crowd-workers on structured text annotation tasks across multiple dimensions.

\textbf{Stage 3: Iterative review.} LLM-generated scores and accompanying rationales underwent multiple rounds of review, drawing on input from the organization's science advisory board and subject-matter consultants. Each review cycle assessed whether scores were consistent with available evidence, whether rationales reflected accurate understanding of each jurisdiction's governance landscape, and whether systematic biases were evident. Scores were revised where reviewers identified errors or unsupported assessments, and revised outputs were re-evaluated until scores and rationales stabilized. This process prioritized substantive accuracy and transparency over formal inter-rater agreement metrics. Independent blind scoring by multiple raters with formal reliability measurement (e.g., Cohen's kappa, Krippendorff's alpha) is planned for the next SRI iteration as the organization's expert network expands.

The LLM-assisted approach represents a methodological innovation with specific advantages and limitations relative to established methods. Compared to the Delphi method \citep{Linstone1975,Hasson2000}, which uses iterative rounds of expert consultation to achieve convergence, LLM-assisted scoring offers greater scalability and perfect reproducibility within a given model version. It sacrifices the genuine diversity of expert perspectives and the deliberative refinement that characterizes Delphi rounds. Compared to the perception-based scoring used in the Worldwide Governance Indicators \citep{Kaufmann2011}, LLM-assisted scoring trades the breadth of multiple independent data sources for consistency and transparency in the scoring process.

The limitations of LLM-assisted scoring must be stated directly. Language models' knowledge is bounded by training data, creating potential recency gaps and geographic biases in coverage. Models may exhibit systematic biases in how they evaluate different jurisdictions. Scoring reproducibility depends on model version; future model updates may yield different results. These limitations are addressed further in Section~\ref{sec:limitations}.

\subsection{Aggregation and Normalization}

Individual category scores are assigned on a 0--100 scale. Overall SRI scores are computed as a weighted arithmetic mean of category scores, with weights as specified in Table~\ref{tab:categories}. Category scores reported in this paper are rounded to the nearest integer for presentation; overall scores are computed from unrounded sub-scores prior to rounding, which may produce minor discrepancies (typically less than 2 points) between the reported overall score and the weighted mean of the rounded category scores.

The choice of arithmetic rather than geometric aggregation has substantive implications. Arithmetic aggregation permits full compensability: a high score in one category can fully offset a low score in another. Geometric aggregation, by contrast, penalizes imbalanced profiles by reducing the overall score when any category is substantially weaker than others \citep{Greco2019,Mazziotta2013}. Given the SRI's finding that Research Environment is universally the strongest category and Professional Readiness the weakest, the choice of aggregation method directly affects how this imbalance is reflected in overall scores. The arithmetic mean used here may overstate readiness by allowing strong research capacity to compensate for weak professional preparedness. This choice is revisited in Section~\ref{sec:limitations} as a limitation, and sensitivity to alternative aggregation methods is reported in Section~\ref{sec:robustness}.

Jurisdictions are classified into five readiness tiers based on overall SRI scores:

\begin{itemize}
  \item \textbf{Well Prepared} (80--100): Comprehensive readiness infrastructure in place
  \item \textbf{Moderately Prepared} (60--79): Substantial infrastructure with identified gaps
  \item \textbf{Partially Prepared} (40--59): Some infrastructure elements but significant deficits
  \item \textbf{Minimally Prepared} (20--39): Limited infrastructure, major gaps across most categories
  \item \textbf{Unprepared} (0--19): Negligible readiness infrastructure
\end{itemize}

\subsection{Robustness and Sensitivity}
\label{sec:robustness}

Saltelli~\citep{Saltelli2007} argues that composite indicators without rigorous sensitivity analysis can be ``constructed to tell any story.'' This critique establishes a standard the SRI must meet. Three robustness checks were conducted.

First, weight sensitivity: category weights were varied by $\pm$5 percentage points across the two most heavily weighted categories (Policy Environment and Professional Readiness, each 20\%). Under all moderate perturbation schemes, rank-order correlations with the baseline remained above $\rho = 0.99$, and tier changes affected at most two jurisdictions. Even under equal weighting (16.7\% per category), the rank correlation held at $\rho = 0.995$ with only one tier change. Only an extreme redistribution (Research Environment and Adaptive Capacity each weighted at 30\%, all others at 10\%) produced substantial tier shifts (10 jurisdictions), though the rank correlation remained high ($\rho = 0.957$). The top two positions (United Kingdom, European Union) and bottom two (Russia, Turkey) were invariant across all weighting schemes tested.

Second, aggregation sensitivity: scores were recomputed using geometric aggregation, which penalizes imbalanced category profiles by reducing overall scores when any single category is substantially weaker. The Spearman rank correlation between arithmetic and geometric rankings was $\rho = 0.982$ ($p < .001$; Kendall's $\tau = 0.906$, $p < .001$), with a mean absolute rank change of 1.13 positions. The maximum rank change observed was 5 positions (Italy, rank 24 to 29), driven by Italy's extremely low Professional Readiness score (5) relative to its Research Environment (55). France moved up 2 positions under geometric aggregation, reflecting its more balanced category profile. The top two and bottom two jurisdictions remained unchanged.

Third, scoring stability: for the 15 jurisdictions with three or more scoring iterations, the mean within-jurisdiction standard deviation across runs was 4.39 points. The most volatile jurisdictions were the Netherlands (SD = 10.99, range 30.4 to 50.3) and the United Arab Emirates (SD = 10.04, range 28.5 to 48.2). The most stable included the United Kingdom (SD = 4.20, range 42.5 to 50.4) and the United States (SD = 4.51, range 38.5 to 50.3), both of which maintained their tier classification across all iterations. This level of volatility is notable and is discussed as a limitation in Section~\ref{sec:limitations}. Given observed within-jurisdiction variability, the tier classifications (Unprepared, Minimally Prepared, Partially Prepared, Moderately Prepared, Well Prepared) represent the appropriate level of interpretive precision for the current methodology. Ordinal rankings within tiers should be treated as indicative rather than definitive; small score differences between closely ranked jurisdictions may fall within the range of scoring noise.

These robustness checks do not eliminate the methodological limitations identified by Saltelli~\citep{Saltelli2007} and Greco et al.~\citep{Greco2019}, but they provide transparency about the sensitivity of results to key design choices.

\section{Results}
\label{sec:results}

\subsection{Overall Rankings}

The headline finding is that no jurisdiction assessed achieves a score above the Partially Prepared tier (40--59). The United Kingdom leads at 49.00, near the top of this range but well below the threshold for Moderately Prepared (60). The global mean is 33.03 (SD = 9.12, median = 35.25). The distribution does not significantly depart from normality (Shapiro-Wilk $W = 0.970$, $p = 0.512$) and shows slight negative skew ($-0.265$) and platykurtosis ($-0.816$), consistent with a moderately compressed distribution concentrated in the lower-middle range.

Eight jurisdictions (25.8\%) reach the Partially Prepared tier (40--59): the United Kingdom (49.00), the European Union (46.75), the United States (45.25), Japan (44.00), Germany (42.30), Australia (41.50), Canada (41.35), and France (40.45). Twenty-one jurisdictions (67.7\%) fall in the Minimally Prepared range (20--39). Two jurisdictions (6.5\%) score below 20: Russia (16.30) and Turkey (14.25). No jurisdiction reaches Moderate (60--79) or Well Prepared (80--100). The Gini coefficient of 0.155 indicates relatively compressed inequality; the global readiness deficit is widespread rather than concentrated.

Table~\ref{tab:rankings} presents the complete rankings.

\begin{table}[htbp]
\centering
\caption{Complete SRI Rankings}
\label{tab:rankings}
\small
\begin{tabular}{@{}rlrc cccccc@{}}
\toprule
Rank & Jurisdiction & Score & Tier & \rotatebox{60}{Policy Env.} & \rotatebox{60}{Inst. Eng.} & \rotatebox{60}{Research Env.} & \rotatebox{60}{Prof. Ready.} & \rotatebox{60}{Pub. Disc.} & \rotatebox{60}{Adapt. Cap.} \\
\midrule
1 & United Kingdom & 49.00 & Partial & 55 & 35 & 70 & 20 & 40 & 75 \\
2 & European Union & 46.75 & Partial & 55 & 25 & 70 & 20 & 40 & 75 \\
3 & United States & 45.25 & Partial & 35 & 45 & 75 & 25 & 40 & 50 \\
4 & Japan & 44.00 & Partial & 48 & 42 & 70 & 18 & 35 & 58 \\
5 & Germany & 42.30 & Partial & 45 & 22 & 70 & 30 & 35 & 62 \\
6 & Australia & 41.50 & Partial & 55 & 19 & 65 & 20 & 30 & 60 \\
7 & Canada & 41.35 & Partial & 45 & 28 & 70 & 18 & 35 & 52 \\
8 & France & 40.45 & Partial & 45 & 38 & 55 & 25 & 30 & 50 \\
9 & Austria & 39.25 & Minimal & 45 & 25 & 55 & 20 & 30 & 60 \\
10 & Spain & 37.70 & Minimal & 45 & 28 & 55 & 15 & 25 & 58 \\
11 & Switzerland & 37.50 & Minimal & 50 & 20 & 60 & 15 & 25 & 55 \\
12 & Norway & 36.50 & Minimal & 45 & 20 & 55 & 15 & 25 & 60 \\
13 & Sweden & 36.25 & Minimal & 45 & 20 & 40 & 15 & 25 & 55 \\
14 & South Korea & 35.75 & Minimal & 45 & 25 & 55 & 20 & 30 & 50 \\
15 & Mexico & 35.50 & Minimal & 62 & 13 & 55 & 15 & 23 & 42 \\
16 & Denmark & 35.25 & Minimal & 35 & 20 & 55 & 15 & 25 & 60 \\
17 & Poland & 33.75 & Minimal & 35 & 25 & 45 & 20 & 30 & 50 \\
18 & Belgium & 33.50 & Minimal & 35 & 15 & 55 & 20 & 25 & 50 \\
19 & Netherlands & 32.35 & Minimal & 15 & 8 & 65 & 18 & 25 & 70 \\
20 & Brazil & 30.00 & Minimal & 35 & 15 & 40 & 20 & 25 & 45 \\
21 & China & 29.35 & Minimal & 28 & 35 & 45 & 15 & 20 & 42 \\
22 & UAE & 28.50 & Minimal & 35 & 15 & 40 & 20 & 15 & 55 \\
23 & Indonesia & 24.65 & Minimal & 28 & 18 & 35 & 15 & 22 & 32 \\
24 & Italy & 23.85 & Minimal & 38 & 5 & 40 & 10 & 5 & 45 \\
25 & Thailand & 23.80 & Minimal & 28 & 12 & 35 & 8 & 15 & 45 \\
26 & Saudi Arabia & 23.50 & Minimal & 28 & 15 & 35 & 10 & 12 & 45 \\
27 & Argentina & 23.30 & Minimal & 30 & 8 & 35 & 15 & 10 & 42 \\
28 & India & 21.50 & Minimal & 18 & 8 & 35 & 12 & 15 & 42 \\
29 & Nigeria & 21.00 & Minimal & 35 & 15 & 30 & 10 & 12 & 25 \\
30 & Russia & 16.30 & Unprep. & 25 & 8 & 20 & 5 & 10 & 18 \\
31 & Turkey & 14.25 & Unprep. & 15 & 5 & 25 & 8 & 12 & 20 \\
\bottomrule
\end{tabular}
\end{table}

\begin{figure}[htbp]
\centering
\includegraphics[width=\textwidth]{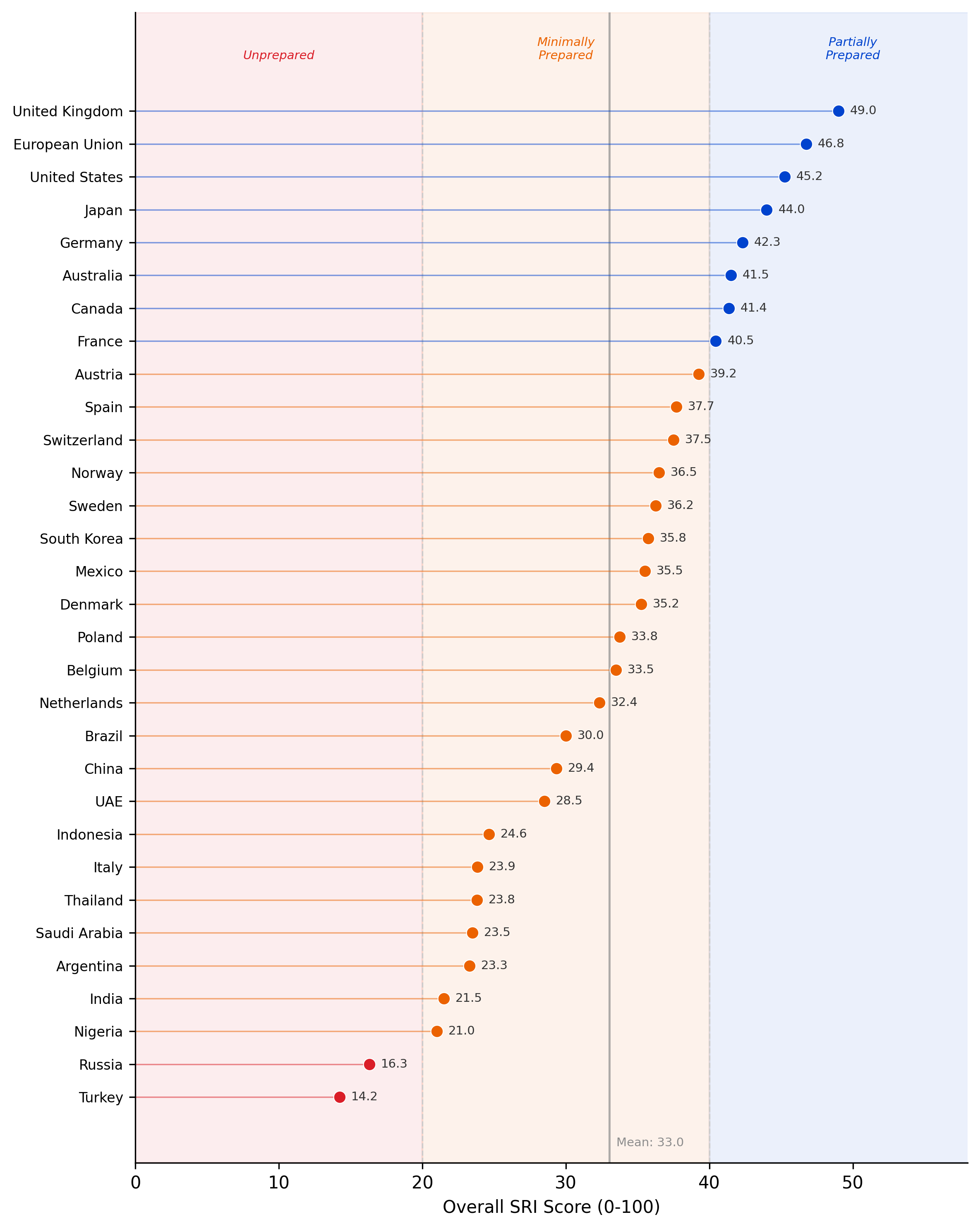}
\caption{SRI Score Distribution. Horizontal lollipop chart showing overall SRI scores for all 31 jurisdictions. Color indicates readiness tier. Vertical dashed lines mark tier boundaries at 20, 40, and 60. The global mean (33.0) is shown as a solid gray line. No jurisdiction exceeds the Partially Prepared tier.}
\label{fig:scores}
\end{figure}

Geographic patterns are visible. North American jurisdictions score highest as a group ($M = 40.70$), followed by Europe ($M = 37.46$) and Asia-Pacific ($M = 31.51$). Middle East and Africa trail at $M = 21.81$. Significant variation exists within regions: Europe spans from 49.00 (UK) to 23.85 (Italy), a 25.15-point range. Asia-Pacific spans from 44.00 (Japan) to 21.50 (India). Some jurisdictions break regional expectations. Mexico (35.50) ranks 15th overall, driven by the highest Policy Environment score in the sample (62), reflecting sustained legislative activity including multiple AI governance bills since 2020 and a proposed constitutional amendment (February 2025) to grant Congress explicit authority over AI regulation. The Netherlands (32.35) ranks 19th despite strong Research Environment (65) and Adaptive Capacity (70) scores, pulled down by the lowest Institutional Engagement score among European jurisdictions (8).

\subsection{Category-Level Analysis}

The most striking category-level finding is the universality and magnitude of the Research Environment and Professional Readiness gap. Table~\ref{tab:catstats} presents descriptive statistics for each category.

\begin{table}[htbp]
\centering
\caption{Category-Level Descriptive Statistics}
\label{tab:catstats}
\small
\begin{tabular}{@{}lcccccccc@{}}
\toprule
Category & Mean & Median & SD & Min & Max & Range & IQR & CV \\
\midrule
Research Environment & 50.16 & 55.0 & 15.14 & 20 & 75 & 55 & 25.0 & 0.30 \\
Adaptive Capacity & 49.94 & 50.0 & 13.70 & 18 & 75 & 57 & 15.5 & 0.27 \\
Policy Environment & 38.16 & 35.0 & 11.88 & 15 & 62 & 47 & 16.0 & 0.31 \\
Public Discourse & 24.06 & 25.0 & 9.61 & 5 & 40 & 35 & 15.0 & 0.40 \\
Institutional Engagement & 20.39 & 20.0 & 10.61 & 5 & 45 & 40 & 11.0 & 0.52 \\
Professional Readiness & 16.52 & 15.0 & 5.44 & 5 & 30 & 25 & 5.0 & 0.33 \\
\bottomrule
\end{tabular}
\end{table}

Research Environment ($M = 50.16$) and Adaptive Capacity ($M = 49.94$) anchor the top of the distribution. Professional Readiness ($M = 16.52$) is the weakest category by a substantial margin. The gap between Research Environment and Professional Readiness scores is positive for all 31 jurisdictions without exception (mean gap = 33.65 points, median = 35.0). A paired-samples $t$-test confirms the gap is statistically significant: $t(30) = 16.03$, $p < .001$. The non-parametric Wilcoxon signed-rank test corroborates: $W = 0.0$, $p < .001$, $r = 1.00$. The effect size is very large (Cohen's $d = 2.88$), placing this among the most consistent patterns in the data.

\begin{figure}[htbp]
\centering
\includegraphics[width=\textwidth]{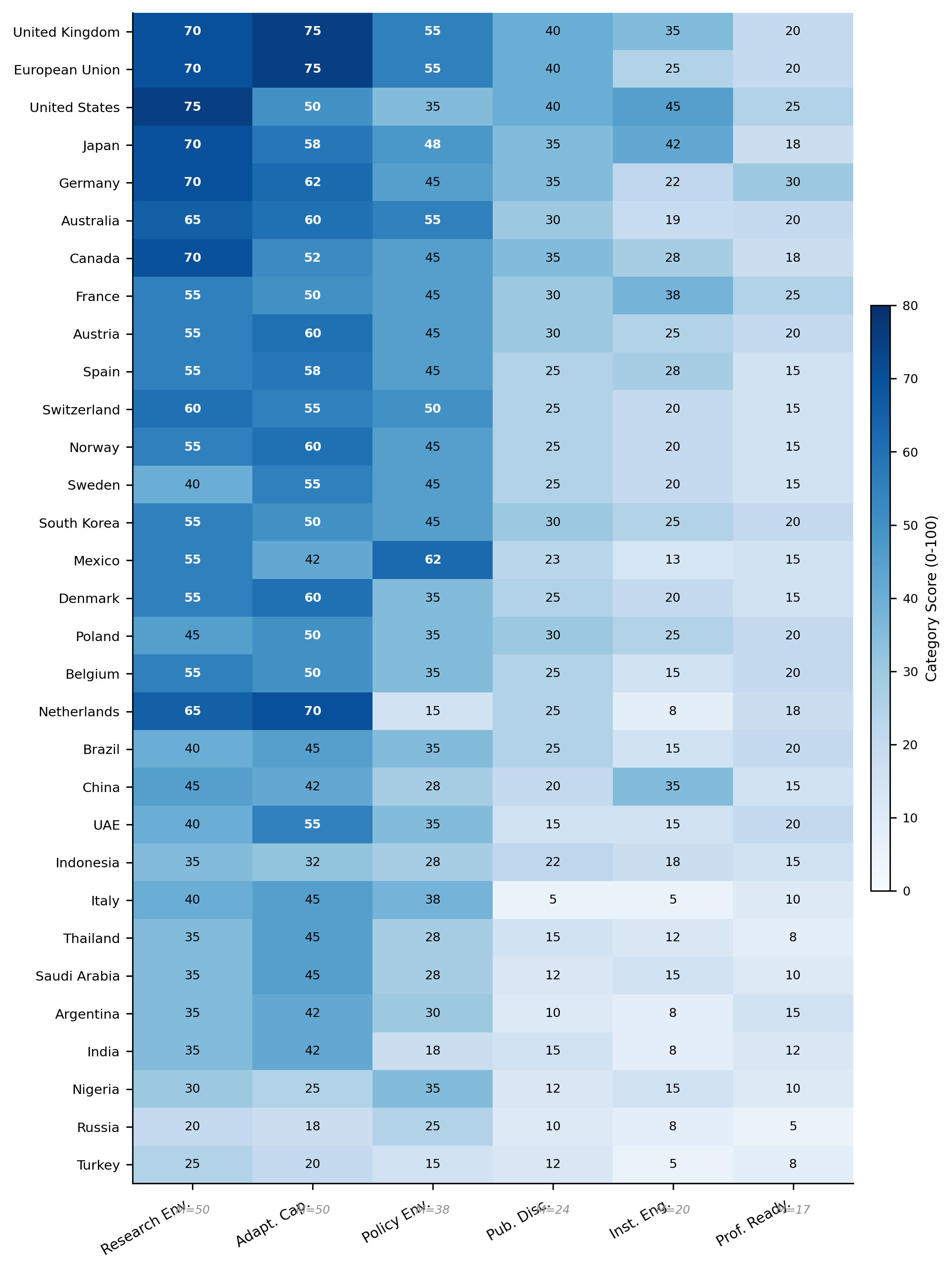}
\caption{Category Performance Heatmap. 31 $\times$ 6 matrix showing jurisdiction scores by category, color-coded from low (dark) to high (light). Jurisdictions sorted by overall SRI score.}
\label{fig:heatmap}
\end{figure}

Institutional Engagement ($M = 20.39$) shows the highest coefficient of variation (CV = 0.52), indicating that this is the category where jurisdictions differ most relative to the mean. The United States (45), Japan (42), and France (38) score well above the category mean, while Italy (5), Turkey (5), and several other jurisdictions score near the floor. Professional Readiness, by contrast, has the narrowest spread (SD = 5.44, range = 25), suggesting a uniformly low ceiling.

\begin{figure}[htbp]
\centering
\includegraphics[width=\textwidth]{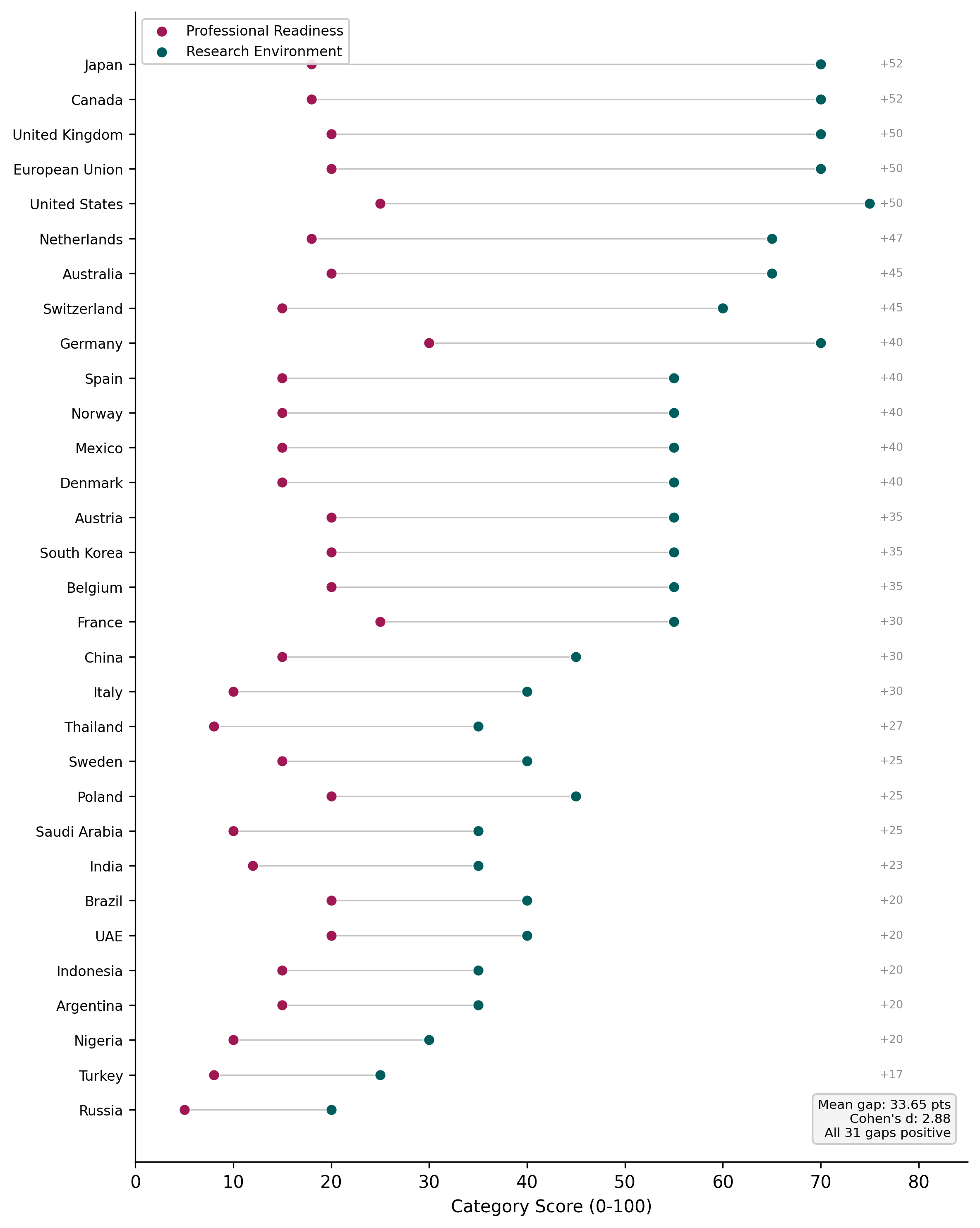}
\caption{Research Environment vs.\ Professional Readiness Gap. Dumbbell chart showing Research Environment and Professional Readiness scores for all 31 jurisdictions, sorted by gap size. Every jurisdiction shows Research Environment exceeding Professional Readiness. Mean gap = 33.65 points, Cohen's $d = 2.88$.}
\label{fig:gap}
\end{figure}

Figure~\ref{fig:gap} visualizes the Research-Professional gap. For every jurisdiction, Research Environment exceeds Professional Readiness, with gaps ranging from 15 points (Russia) to 52 points (Japan and Canada).

Cross-category correlations are uniformly positive and predominantly strong. All 15 pairwise Pearson correlations are statistically significant ($p < .05$), with 11 of 15 significant at $p < .001$; Spearman rank correlations yield a similar pattern. The strongest correlation is between Research Environment and Public Discourse ($r = .876$, $p < .001$), suggesting that jurisdictions with active consciousness research tend to generate higher-quality public conversation. Institutional Engagement correlates strongly with Public Discourse ($r = .767$, $p < .001$), indicating that institutional activity and informed debate co-occur. The weakest pairwise correlation is between Institutional Engagement and Adaptive Capacity ($r = .423$, $p < .05$), suggesting these represent relatively distinct dimensions.

An exploratory principal component analysis of the six categories reveals that the first component explains 70.7\% of total variance, with all six categories loading in the same direction. This suggests a dominant general ``readiness factor'': jurisdictions that score high on one category tend to score high on others. The second component (10.4\% of variance) distinguishes between jurisdictions strong in Policy Environment and Adaptive Capacity versus those strong in Institutional Engagement and Professional Readiness, capturing a contrast between governance infrastructure and institutional activity.

Several jurisdictions exhibit distinctive profiles. The Netherlands (overall: 32.35) combines the third-highest Adaptive Capacity score (70) and strong Research Environment (65) with the lowest Institutional Engagement score among all European jurisdictions (8), yielding the most imbalanced profile in the sample (within-jurisdiction CV = 0.80, max-min spread = 62 points). Italy (23.85) shows a similarly extreme profile: moderate Policy Environment (38) and Adaptive Capacity (45) alongside the lowest Institutional Engagement (5) and Public Discourse (5) scores in the dataset. China (29.35) presents a different pattern: its Institutional Engagement score (35, fifth-highest) reflects mandated AI ethics committees and a comprehensive regulatory apparatus, though these institutions prioritize security, economic competitiveness, and social stability rather than inquiry into AI consciousness specifically. China's score on this category captures institutional activity that could be redirected toward sentience governance, not current engagement with sentience questions.

\subsection{Comparative Analysis with Existing Indices}

A central question for any new composite index is whether it measures something distinct from existing instruments. If the SRI simply reproduced established AI readiness rankings, it would add measurement without adding insight. We would expect strong Research Environment and Adaptive Capacity (the categories most aligned with general governance and research capacity) to dominate the rankings. These categories are indeed the strongest. But the categories that most differentiate the SRI from existing indices, Institutional Engagement, Professional Readiness, and Public Discourse, are precisely where scores are lowest. The mean of the three ``novel'' categories (Institutional Engagement, Professional Readiness, Public Discourse) is 20.32, compared to 46.09 for the three categories with the most overlap with general governance indices (Policy Environment, Research Environment, Adaptive Capacity). This 25.76-point differential indicates that sentience-specific readiness substantially lags general governance capacity, supporting the SRI's premise that a distinct measurement dimension is needed.

Jurisdiction-level divergences reinforce this point. The Netherlands ranks 7th on the Oxford Insights Government AI Readiness Index \citep{OxfordInsights2024} but 19th on the SRI, a 12-position drop driven by the lowest Institutional Engagement score among European jurisdictions (8) and weak Policy Environment (15). The United States ranks 1st on Oxford Insights but 3rd on the SRI, with its lead eroded by middling Policy Environment (35) and the same Professional Readiness deficit seen globally. Conversely, Mexico ranks 15th on the SRI but falls well outside the top 50 on most general AI readiness indices, buoyed by the highest Policy Environment score in the sample (62). These divergences are not random; they reflect the fact that general AI governance capacity and sentience-specific readiness draw on partially overlapping but distinct institutional foundations. A jurisdiction can be a global AI leader while having negligible infrastructure for addressing AI moral status.

\subsection{Regional and Governance-Type Patterns}

Regional differences in SRI scores are statistically significant (Kruskal-Wallis $H(4) = 12.95$, $p = .012$, $\eta^2 = 0.36$). North America ($M = 40.70$, $n = 3$) and Europe ($M = 37.46$, $n = 14$) score highest, followed by Asia-Pacific ($M = 31.51$, $n = 7$). The Middle East and Africa group ($M = 21.81$, $n = 4$) and Latin America ($M = 26.65$, $n = 2$) score lower. These patterns broadly track economic development and research infrastructure, but regional means mask substantial within-region variation. Europe spans 25.15 points from the UK (49.00) to Italy (23.85). Asia-Pacific spans 22.50 points from Japan (44.00) to India (21.50).

Governance type yields the clearest pattern in the data. Using the Economist Intelligence Unit's Democracy Index \citep{EIU2024} classification, democracies ($n = 24$) score significantly higher than hybrid or authoritarian regimes ($n = 7$) on overall SRI: $M = 36.13$ vs.\ $M = 22.39$, Mann-Whitney $U = 156.0$, $p < .001$, rank-biserial $r = -0.857$. The effect holds across all six categories, with the largest absolute differences in Research Environment (+22.35 points), Adaptive Capacity (+18.37), and Public Discourse (+13.37). These three categories, representing the dimensions most dependent on open inquiry, institutional flexibility, and freedom of expression, account for the majority of the governance-type differential. Institutional Engagement (+6.96) and Professional Readiness (+7.31) show the smallest gaps, reflecting that these categories are weak across governance types, though for different reasons: institutional engagement with sentience questions is nascent everywhere, while professional preparedness is near-universally absent.

\begin{figure}[htbp]
\centering
\includegraphics[width=\textwidth]{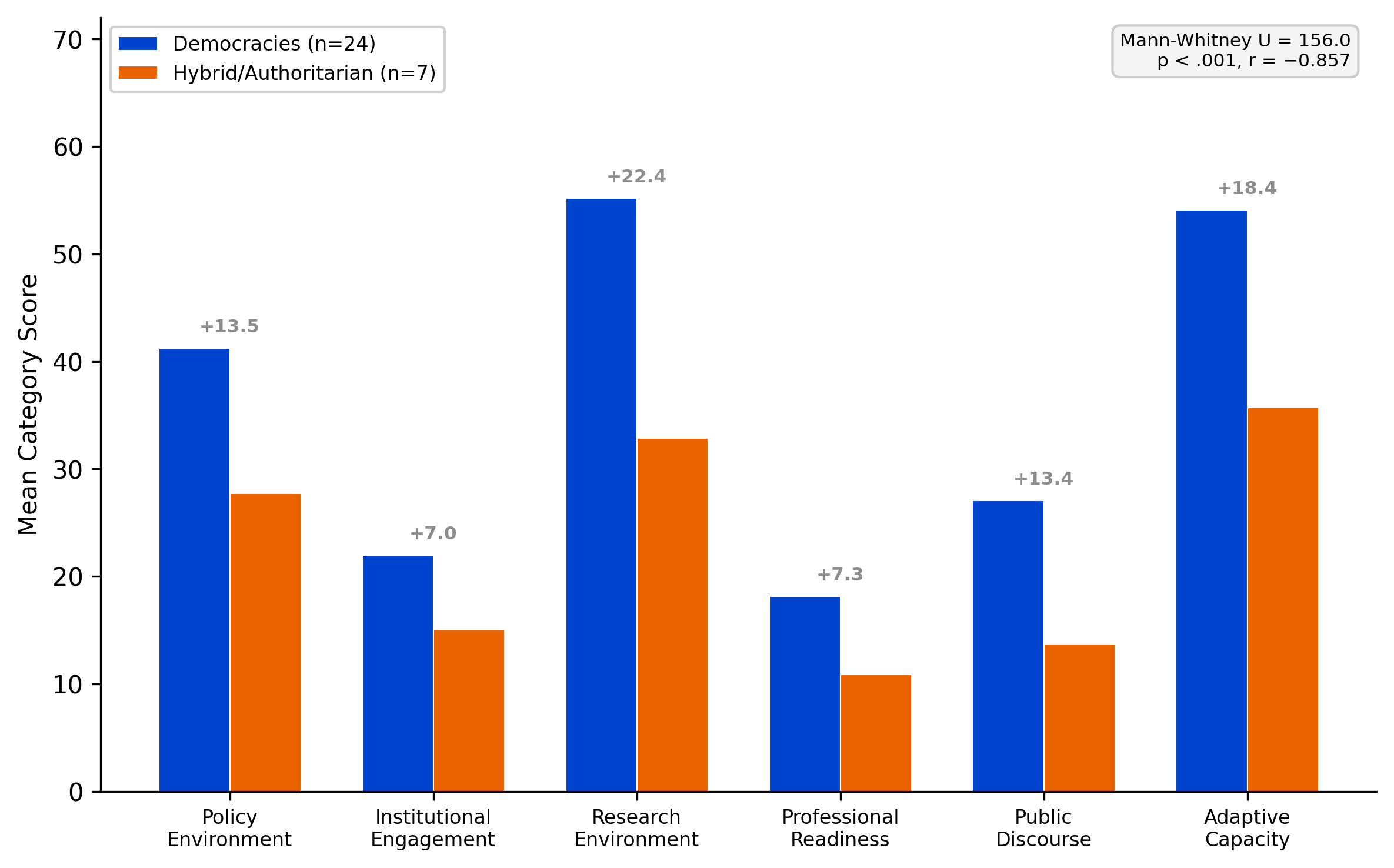}
\caption{Category-level SRI score comparison between democracies ($n = 24$) and hybrid/authoritarian regimes ($n = 7$). All six categories show higher mean scores for democracies, with the largest differentials in Research Environment (+22.35), Adaptive Capacity (+18.37), and Public Discourse (+13.37). Mann-Whitney $U = 156.0$, $p < .001$, rank-biserial $r = -0.857$.}
\label{fig:governance}
\end{figure}

Income classification, based on World Bank~\citep{WorldBank2024} groupings, is also significantly associated with SRI scores (Kruskal-Wallis $H(2) = 14.31$, $p < .001$, $\eta^2 = 0.44$). High-income jurisdictions ($n = 21$) average 37.35, upper-middle-income ($n = 7$) average 24.64, and lower-middle-income ($n = 3$) average 22.38. The gap between high-income and upper-middle-income jurisdictions (12.70 points) is substantially larger than between upper-middle and lower-middle (2.26 points), suggesting diminishing marginal association between income and sentience readiness. Mexico (35.50) is the strongest-performing upper-middle-income jurisdiction, scoring above seven high-income jurisdictions and outperforming income expectations primarily through its unusually strong Policy Environment (62).

\section{Discussion}
\label{sec:discussion}

\subsection{The Global Readiness Deficit}

The SRI's central finding, that no jurisdiction exceeds the Partially Prepared tier, is not surprising but it is significant. AI sentience governance is a novel challenge, and societies cannot be expected to have built comprehensive infrastructure for a problem that has only recently moved from philosophical speculation to scientific investigation. The SRI does not indict any jurisdiction for failing a test it was never designed to pass. It provides a diagnostic baseline.

Historical parallels are instructive. Societal readiness for climate change, bioethics, and nuclear governance all showed similar early-stage deficits before institutional infrastructure developed. The Asilomar conference on recombinant DNA \citep{Berg1975} preceded by decades the regulatory frameworks that now govern genetic engineering. The Intergovernmental Panel on Climate Change was established in 1988 \citep{Agrawala1998}, but effective climate governance is still developing nearly four decades later. In each case, the critical step was recognizing that readiness needed to be assessed, gaps identified, and capacity deliberately built. The SRI represents that step for AI sentience governance.

The more consequential question is not whether societies are currently ready (they are not) but whether they are building readiness. The SRI, applied longitudinally, can track whether jurisdictions are developing the institutional, professional, and cultural infrastructure needed to respond to credible sentience claims. The baseline established here makes future progress or stagnation measurable.

\subsection{The Research-Practice Gap}

The universal finding that Research Environment is the strongest category and Professional Readiness the weakest demands interpretation. Consciousness science is an established research field: journals such as \textit{Journal of Consciousness Studies}, \textit{Neuroscience of Consciousness}, and \textit{Journal of Artificial Intelligence and Consciousness} have published for decades. Dedicated research centers exist at major universities. Funding streams, while modest by the standards of AI safety research, are established. The SRI's Research Environment scores reflect this accumulated institutional capacity.

Professional Readiness, by contrast, measures something that barely exists, even as the practical need for it grows. Therapists are already encountering patients who form deep emotional bonds with AI companions and grieve when those systems are shut down or altered; the clinical literature offers little guidance on how to assess these attachments or when they might raise welfare-relevant questions about the systems involved. Teachers report students who attribute feelings and preferences to classroom AI tools, and no educational framework helps distinguish productive anthropomorphism from confusion that could have ethical consequences. The emerging ``grieftech'' industry, which uses AI to simulate deceased loved ones, raises questions about the moral status of interactive digital entities that no medical or counseling professional body has begun to address. Media coverage of AI sentience remains largely sensationalized or dismissive, rarely informed by the scientific literature reviewed in Section~\ref{sec:background}. Technology companies have developed AI safety and ethics frameworks, but these address risks to humans from AI, not the question of whether AI systems themselves might have interests worth protecting.

This gap has a structural explanation. Knowledge about consciousness has been produced within specialized academic communities. The professions that would need to act on that knowledge, legal, healthcare, media, and technology, have had no institutional incentive to engage with it. The SRI identifies this research-practice gap as a priority for policy intervention. Susskind and Susskind~\citep{Susskind2015} documented how technology transforms expert professions, but their analysis focused on service delivery disruption, not on professions' capacity to adjudicate novel moral status questions. Topol~\citep{Topol2019} identified a similar gap between AI promise and clinical readiness in healthcare. The sentience readiness gap the SRI identifies extends this pattern into new territory: the professional knowledge gap is not about competence with tools but about preparedness for entities that might warrant moral consideration.

\subsection{Engaging Objections}

Three objections to the SRI merit direct response.

\textbf{Objection 1: Premature to measure readiness for something that may never happen.} The anticipatory governance framework \citep{Guston2014} and the Collingridge~\citep{Collingridge1980} dilemma, discussed in Section~2.2, provide the theoretical response: governance capacity must be built before crises, and waiting for certainty ensures that capacity arrives too late. The empirical grounding is not speculative: as reviewed in Section~1.1, specialists assign non-negligible probabilities to digital minds within the next decade, comparable to those that justify investment in pandemic preparedness and climate adaptation. The SRI also measures capacities with value beyond the sentience question. Adaptive Capacity and Research Environment, the two strongest categories, reflect institutional flexibility and scientific infrastructure that serve any novel technological challenge. The sentience-specific categories (Institutional Engagement, Professional Readiness, Public Discourse) are more narrowly targeted, but even these build transferable capacities: engaging ethically with claims about non-human minds, preparing professionals for morally complex AI scenarios, and fostering informed public deliberation. These capacities strengthen responsible AI governance whether or not AI sentience is confirmed.

\textbf{Objection 2: The SRI embeds contested value judgments.} This is true of every composite index. Saltelli~\citep{Saltelli2007} and Greco et al.~\citep{Greco2019} establish that composite indicators necessarily reflect normative choices in variable selection, weighting, and aggregation. The SRI's response is not to deny this but to make its commitments transparent. The category weights are stated explicitly, their rationale is described, and sensitivity analysis (Section~\ref{sec:robustness}) reports how results change under alternative weightings. The SRI's normative commitment is minimal: societies should be prepared to respond appropriately to credible sentience claims. This is consistent with Birch's~\citep{Birch2024} precautionary framework, which requires only that we take seriously the possibility of sentience, not that we resolve the question. Floridi~\citep{Floridi2023} articulates ethical principles for AI that could inform which aspects of readiness matter most. Coeckelbergh~\citep{Coeckelbergh2022} raises political philosophy dimensions, including democracy and epistemic agency, that the SRI's weighting implicitly addresses. Jasanoff~\citep{Jasanoff2016} argues technology governance requires democratic deliberation, not just expert assessment. The SRI acknowledges these perspectives and treats its own framework as one defensible operationalization, not the only possible one.

\textbf{Objection 3: LLM-assisted scoring is not genuine expert assessment.} The concern is legitimate. LLM-based evaluation is a novel methodology that has not been validated across all domains. Two responses are warranted. First, the empirical evidence for LLM-as-judge methods is growing: Zheng et al.~\citep{Zheng2023} demonstrate over 80\% agreement with human preferences, and Gilardi et al.~\citep{Gilardi2023} show LLMs outperforming crowd-workers in structured assessment tasks. The SRI incorporates iterative review with advisory board input to catch errors and biases in LLM-generated scores (Section~3.3), and formal inter-rater reliability testing is planned for future iterations. Second, the SRI's LLM-assisted approach should be understood not as a replacement for expert assessment but as a scalable first step. Traditional Delphi methods \citep{Linstone1975} remain the gold standard for expert consensus, and Delphi validation is a priority for future work. The LLM-assisted approach offers reproducibility and scalability that traditional methods cannot match at this scope.

\subsection{Implications for AI Governance}

The SRI reveals a structural blind spot in global AI governance. The EU AI Act \citep{EUAIAct2024}, widely considered the most comprehensive AI regulation enacted to date, takes an explicitly human-centric approach with no provisions addressing AI moral status. The OECD AI Principles \citep{OECD2024} address safety, fairness, transparency, and accountability but not consciousness. UNESCO's Recommendation on the Ethics of AI \citep{UNESCO2021} covers human rights, inclusion, and environmental sustainability but does not contemplate AI systems as potential moral patients. This pattern is not accidental. Current AI governance frameworks were designed to manage AI as a tool, and they accomplish that task with varying degrees of success. They were not designed to address the possibility that AI systems might themselves have morally relevant properties. The SRI measures the gap between existing governance infrastructure and what would be needed for this different kind of challenge.

The governance proposals reviewed in Section~\ref{sec:background} \citep{Birch2024,ButlinLappas2025,Leibo2025,Long2024} converge on a common theme: institutional capacity for responding to sentience claims should be built proactively rather than reactively. The SRI's results indicate where that capacity is weakest. Professional Readiness and Institutional Engagement, the two categories most directly aligned with these proposals, are also the two lowest-scoring categories globally. The gap between what the literature calls for and what exists is not abstract; it is quantified in the SRI's category-level findings.

The public perception dimension is equally important. As discussed in Section~2.3, public beliefs about AI sentience will shape policy regardless of expert consensus \citep{Caviola2025}, and institutional trust mediates the translation of concern into regulation support \citep{Bullock2025}. Effective sentience governance will require not only good policies but public confidence in the institutions implementing them.

The SRI is not a blueprint for sentience-ready governance. It is a diagnostic tool that identifies where specific capacities are strong, where they are weak, and where jurisdictions might prioritize investment. Its value lies in making the readiness gap visible, measurable, and tractable.

\subsection{Epistemic Humility and Scope}

The SRI's claims are bounded by its design. It measures institutional, professional, and cultural capacities; it does not assess the likelihood of AI sentience, the desirability of preparing for it, or the adequacy of any particular policy response. The precautionary logic is conditional, and the condition may never be met. A responsible reading of the SRI is therefore prospective: the index identifies capacity gaps that would matter if certain scientific developments occur, not capacity gaps that necessarily matter now. The asymmetry noted in Section~5.3 bears restating in its sharpest form: the cost of building capacity that proves unnecessary is low; the cost of needing capacity that was never built could be high.

\section{Limitations}
\label{sec:limitations}

Several limitations of the preliminary SRI must be acknowledged.

\textbf{Methodological limitations.} The SRI is a single-wave assessment reflecting conditions as of December 2025. Sentience readiness is likely to evolve as AI governance matures, public awareness shifts, and scientific understanding develops. Longitudinal tracking is needed to capture this evolution. The weighted arithmetic aggregation permits full compensability between categories, meaning that strong performance in one area can mask weakness in another. Alternative aggregation methods, particularly geometric aggregation \citep{Mazziotta2013}, would penalize such imbalances and might yield different rankings. The weighting scheme, while defended on substantive grounds, reflects normative choices that others might reasonably make differently. Seth and McGillivray~\citep{SethMcGillivray2018} provide methods for testing ranking robustness under alternative weights that should be applied more extensively in future iterations.

\textbf{Scope limitations.} The 31 jurisdictions assessed do not constitute a comprehensive global sample. Many countries with developing AI ecosystems are not included. The national-level assessment misses subnational variation, which can be substantial in federal systems or in countries where AI governance capacity is concentrated in specific regions or cities. The assessment is dated December 2025; given the pace of AI governance development, some findings may already reflect outdated conditions.

\textbf{Conceptual limitations.} Measuring ``readiness'' for an uncertain phenomenon is inherently contestable. The SRI operationalizes one defensible interpretation of what readiness means in this context, but alternative operationalizations could emphasize different capacities or weight them differently. The concept of readiness also carries an implicit assumption that preparedness is desirable, which some might contest on grounds that preparing for AI sentience could normalize premature sentience attributions.

\textbf{LLM-specific limitations.} Language models' knowledge is bounded by training data, creating potential recency gaps and geographic biases in coverage. Models from different developers may exhibit systematically different evaluation patterns. Scoring reproducibility depends on model version; future model updates could yield meaningfully different results. The extent to which LLMs can genuinely assess complex governance phenomena, as opposed to reproducing patterns from their training data, remains an open question.

\textbf{Validation limitations.} The preliminary version relies on iterative review rather than formal inter-rater reliability testing. No independent blind scoring was conducted, meaning score reliability cannot be quantified using standard agreement metrics such as Cohen's~\citep{Cohen1960} kappa or Krippendorff's~\citep{Krippendorff2004} alpha. Formal validation with independent raters, and ideally a full Delphi process, is planned for future iterations.

\section{Conclusion}
\label{sec:conclusion}

This paper has introduced the preliminary Sentience Readiness Index, the first composite index measuring national-level preparedness for the possibility of artificial sentience. The SRI assesses 31 jurisdictions across six weighted categories and finds a consistent pattern: no society is well prepared. The United Kingdom leads at 49 out of 100, and the majority of jurisdictions fall in the Minimally Prepared range. Research Environment is universally the strongest category, reflecting decades of academic work on consciousness science. Professional Readiness is universally the weakest, indicating that the healthcare, legal, media, and technology professions have not begun to prepare for scenarios involving AI moral status.

These findings are not a cause for alarm. AI sentience governance is a novel challenge, and the absence of established infrastructure is expected. What matters is whether societies begin building that infrastructure now, while the question of AI consciousness remains open and institutional frameworks remain flexible. The Collingridge~\citep{Collingridge1980} dilemma warns that delayed governance is ineffective governance. The precautionary framework \citep{Birch2024} provides the ethical foundation for acting under uncertainty. The SRI provides the measurement infrastructure for tracking whether action is being taken.

The SRI establishes a baseline. Future iterations can track longitudinal change, expand to additional jurisdictions, and incorporate refined scoring methodologies as the field develops. The index is not the final word on sentience readiness. It is, to our knowledge, the first systematic attempt to make sentience readiness visible, measurable, and subject to informed policy action.

Whether AI systems will ever be conscious remains an open scientific question. Whether societies are prepared for the possibility that they might be is an institutional question that the SRI can help answer. The gap between these two questions, one scientific and perhaps unresolvable, the other institutional and clearly actionable, is where the SRI makes its contribution.

\begin{ack}
\textbf{Funding}: None.

\textbf{Competing interests}: The author is the Executive Director of the organization that developed the SRI. No additional financial interests.

\textbf{Ethics approval}: Not applicable. This study did not involve human or animal subjects.

\textbf{AI disclosure}: Large language models were used in the SRI scoring methodology, as described in detail in Section~3.3. LLMs were also used to assist with copy editing of the manuscript text. All content reflects the author's original analysis and conclusions.

\textbf{Data availability}: SRI scores, category-level data, and scoring rubrics are publicly available at \url{https://harderproblem.org/sri/api/} and archived at \url{https://zenodo.org/records/18780234} (DOI: 10.5281/zenodo.18780233).
\end{ack}


\appendix


\end{document}